\documentclass[aps,pre,nopacs,raggedbottom, nobalancelastpage,
amssymb,twocolumn,groupedaddress,longbibliography]{revtex4}
\usepackage{graphicx}
\usepackage{amsmath}
\usepackage{amsfonts}
\usepackage{amssymb}
\usepackage{epsfig,libertine}
\usepackage[libertine]{newtxmath}
\usepackage{color}
\usepackage{dsfont, hyperref, xcolor}
\usepackage{comment}
\usepackage{enumerate}
\usepackage{mathtools}
\usepackage{physics}
\usepackage{natbib}
\usepackage{svg}
\usepackage{array, dcolumn}
\begin{document}

\title{Keldysh field theory approach to electric and thermoelectric transport in quantum dots}

\author{Marco Uguccioni, Luca Dell'Anna}
\affiliation{Dipartimento di Fisica e Astronomia e Sezione INFN, Universit\`{a} degli Studi di Padova, via Marzolo 8, 35131 Padova, Italy}

\date{\today}

\begin{abstract}
We compute the current and the noise power matrix in a quantum dot connected to two metallic reservoirs by using the Keldysh field theory approach, a non-equilibrium quantum field theory language in the functional integral formalism. We first show how this technique allows us to recover rapidly and straightforwardly well-known results in literature, such as the Meir-Wingreen formula for the average current, resulting extremely effective in dealing with quantum transport problem. We then discuss in detail the electric and thermoelectric properties due to transport of electrons in the case of a single-level and two-level non-interacting quantum dot. In particular, we derive the optimal conditions for maximizing the thermoelectric current, finding an upper limit for the thermoelectric coefficient. 
Moreover, in the two-level system we show   
that the zero-temperature linear conductance 
drops rapidly to zero by a symmetrical removal of the degeneracy at the Fermi energy.
\end{abstract}

\maketitle

\section{Introduction}

The field of \emph{electronic transport} in nanoscale devices is undergoing rapid evolution, driven by advancements in fabrication techniques and fueled by the potential applications in emerging fields such as spintronics and quantum information processing {\color{black}\cite{bib:1new,bib:2new,bib:3new, bib:4new}}. Among the myriad of nanostructures, \emph{quantum dot} (QD) systems occupy a central position, manifesting in various physical forms, including semiconducting heterostructures, metallic nanoparticles, carbon nanotubes, and molecular assemblies coupled to metallic electrodes {\color{black}\cite{bib:5new,bib:6new}}. Despite their diverse realizations, these devices often lend themselves to theoretical descriptions characterized by simple "universal-like" models, governed by a handful of parameters \cite{bib:1new}. Beyond their practical applications, QD systems serve as invaluable platforms for exploring the interplay between electronic correlations and transport phenomena under non-equilibrium conditions.

The investigation of electron transport in semiconducting QDs dates back to the early 1990s, marked by the seminal observation of phenomena like \emph{Coulomb blockade} \cite{bib:1}. It swiftly became evident that QDs offer a unique arena for probing fundamental electronic correlation phenomena such as the \emph{Kondo effect}, as initially proposed in theoretical predictions \cite{bib:2, bib:3}. These predictions found initial experimental validation in metallic nanoscale junctions containing magnetic impurities \cite{bib:4}. However, a definitive breakthrough occurred with the observation of the Kondo effect in semiconducting QDs \cite{bib:5,bib:6}. Notably, these QD devices provide the means to control relevant parameters, facilitating direct comparisons with theoretical predictions. Subsequently, the manifestation of Kondo correlations in electronic transport has been demonstrated in various physical incarnations of QDs, including carbon nanotubes \cite{bib:7} and large molecules such as fullerenes \cite{bib:8}.

In addition to electric transport, there is growing interest in investigating \emph{thermoelectric transport} phenomena in QD-based systems, which contain complementary information about the device. Thermoelectric effects enable the conversion of temperature gradients into electric voltages (\emph{Seebeck effect}), offering opportunities for efficient energy harvesting. By exploring thermoelectric properties in QD junctions, researchers aim to elucidate the mechanisms governing heat and charge transport at the nanoscale, with implications for the development of advanced thermoelectric materials and devices. We refer to \cite{bib:9} for a review of experimental studies on QD thermoelectric properties.

In the pursuit of understanding and manipulating electronic and thermoelectric transport in QD-based junctions, the computation of physical quantities such as currents and noises play a crucial role. Indeed, the accurate calculation of thermal and electric charge current flow provides insights into the fundamental transport mechanisms, while characterization of the noise power in QD junctions offers valuable information about the coherence properties, quantum fluctuations, and dissipation processes occurring within these nanoscale systems.

From a theoretical point of view, quantum transport in microelectronic devices has been mainly addressed by {\color{black} three} different approaches {\color{black} \cite{bib:ryndyk}}. One is based on the scattering picture first introduced by Landauer \cite{bib:landauer}, and generalized by Buttiker \cite{bib:buttiker}, in which the transport properties are expressed in terms of transmission and reflection scattering amplitudes. This approach is generally used in a phenomenological way by replacing the non-interacting device with a simple scattering structure. {\color{black} A second theoretical framework, which allows to deal with interactions, is the quantum master equation (QME) technique \cite{bib:petruccione}, based on the density matrix in the basis of the many-body states of the isolated system. This framework is used in the case of very weak system-to-leads coupling, it gives a fairly complete description of sequential tunneling, of the main features of Coulomb blockade and even of the  Kondo physics for certain temperatures. The third,} more powerful and ab-initio approach is provided by \emph{non-equilibrium Green's functions} (NEGF) technique \cite{bib:jauho}, which can be extended to interacting systems. The latter approach, being part of a \emph{quantum field theory} (QFT) language, has been mainly developed in the \emph{second quantization} operator language, which could be perceived as too technical and cumbersome to learn. Following this line of reasoning, in this paper we take a different direction by working with a \emph{functional integrals} QFT formalism, also known as \emph{Keldysh field theory} \cite{bib:keldysh, bib:kamenev, bib:diehl} (KFT), which offers a more transparent, systematic and efficient means to obtain physical quantities straightforwardly, enabling a comprehensive understanding of transport phenomena in non-equilibrium conditions.

The paper is organized as follows: in Sec. \ref{sec:model}, we present
the system and the interesting physical observables, namely the average current and the noise power. In Sec. \ref{sec:KFT} we present the general KFT formalism used to deal with quantum transport problems, which allows to recover rapidly and straightforwardly well-known results in literature. In Sec. \ref{bib:single}, we analyze in detail the current, with considerations on the conductance and the thermoelectric coefficient, and the noise power matrix in a single-level non-interacting QD. We compare how the current properties change when one considers a two-level non-interacting QD in Sec. \ref{bib:two}. We give some conclusions in Sec. \ref{sec:conclusion}.

\section{Model}
\label{sec:model}
QD are theoretically described as point-like, or zero dimensional (0D), entities with discrete levels. Here we consider a generic interacting QD coupled to two metallic leads (labeled as $a=L, R$), which can be modeled as free electron systems. The total system is described by the Hamiltonian $\hat{H}=\hat{H}_{leads}+\hat{H}_{dot}+\hat{H}_T$, where
 \begin{equation}
    \hat{H}_{leads}=\sum_k\sum_\sigma \bigg[\omega_{kL}\hat{c}_{k\sigma L}^\dagger\hat{c}_{k\sigma L} +\omega_{kR}\hat{c}_{k\sigma R}^\dagger\hat{c}_{k\sigma R}\bigg],
\end{equation}
\begin{equation}
\label{eq:Hdot}
    \hat{H}_{dot}=\sum_\sigma\sum_n \varepsilon_n\hat{d}^\dagger_{n\sigma}\hat{d}_{n\sigma}+\hat{V}(\hat{d}^\dagger_{n\sigma},\hat{d}_{n\sigma}),
\end{equation}
\begin{equation}
    \hat{H}_{T}=\sum_{a=L,R}\sum_k \sum_n\sum_\sigma\bigg[W_{ka,n}\hat{c}^\dagger_{k\sigma a}\hat{d}_{n\sigma}+W^*_{ka,n}\hat{d}_{n\sigma}^\dagger\hat{c}_{k\sigma a}\bigg],
\end{equation}
Here, the operators $\hat{c}^\dagger_{k\sigma a}$ ($\hat{c}_{k\sigma a}$)  create (annihilate) electrons with momentum $k$, spin $\sigma$, and dispersion relation $\omega_{ka}$ in the corresponding lead $a=L,R$. Similarly, the operators $\hat{d}^\dagger_{n\sigma}$ ($\hat{d}_{n\sigma}$) create (annihilate) electrons with spin $\sigma$ in the discrete states $n=0,1,2,\dots$ of the QD, with non-interacting energy $\varepsilon_n$. Finally, $W_{ka,n}$ are tunneling matrix elements between the leads and the dot. Notice that, in the Hamiltonian $\hat{H}_{dot}$ of the isolated dot, we included arbitrary interactions $\hat{V}$. In particular, if one considers a single discrete level with a local Coulomb interaction for the electrons in the level of the type $\hat{V}=U\hat{d}^\dagger_{\uparrow}\hat{d}_{\uparrow}\hat{d}^\dagger_{\downarrow}\hat{d}_{\downarrow}$, the total Hamiltonian above would become the well-known \emph{Anderson impurity model} \cite{bib:anderson}, which is usually used in the analysis of transport in QDs and their peculiar effects, such as the Kondo effect \cite{bib:kondo} or the Coulomb Blockade \cite{bib:coulomb}. In this work, we neglect the Coulomb interaction between electrons in the dot, even though the derivation in the section below is completely general and holds even in the presence of interactions in the dot. This assumption is reasonable when at least one of the characteristic energies of the problem (i.e. dot energy, temperature, voltage, coupling between dot and reservoirs) is large in comparison to the Coulomb energy $U$.

We consider the presence of an external voltage $V$ applied to the system, which creates a current due to a bias in the chemical potentials $\mu_a$ of the leads, $eV=\mu_L-\mu_R$, and we choose the energy origin at the right chemical potential, namely $\mu_R=0$. A convenient way to take into account the effects of the external voltage $V$ is to perform a \emph{gauge transformation} on the fermionic operators, to move the voltage to time-dependent hoppings
\begin{equation}
\label{eq:T_gauge}
W_{ka,n}\to W_{ka,n}(t)=W_{ka,n}e^{-i\mu_at}.
\end{equation}
We thus define the current operator from the contact $a=L,R$ to the central region as
\begin{equation}
\begin{split}
\label{eq:Jtransport}
\hat{J}_a(t)&=-e\frac{d}{dt}\hat{N}_a(t)= -e\frac{d}{dt}\sum_k\sum_\sigma \hat{c}^\dagger_{k\sigma a }(t)\hat{c}_{k\sigma a}(t)\\&=-ie\sum_\sigma\big[\hat{H}(t),\hat{c}^\dagger_{k\sigma a}(t)\hat{c}_{k\sigma a}(t)\big]\\&=ie\sum_k\sum_n\sum_\sigma \bigg[W_{ka,n}(t)\hat{c}^\dagger_{k\sigma a}(t)\hat{d}_{n\sigma}(t)+\\&-W^*_{ka,n}(t)\hat{d}_{n\sigma}^\dagger(t)\hat{c}_{k\sigma a}(t)\bigg],
\end{split}
\end{equation}
where in the second line we used the \emph{Heisenberg equations of motion}, while in the third line we performed the commutation relation explicitly. 

The current is not the only quantity which can help to understand the transport properties of a system. Another interesting magnitude that can be calculated using the non-equilibrium formalism below (especially using functional integrals) is the \emph{noise} due to \emph{current fluctuations}. The noise in a mesoscopic system like the one we are studying can be either a consequence of the thermal movements of electrons (\emph{thermal noise} or \emph{Johnson-Nyquist noise}) or the discreteness of the charge carriers (\emph{shot noise}), and its measurement can provide information not available in usual conductance experiments. The noise is characterized by the \emph{noise spectral density} (or \emph{noise power}) matrix $S_{ab}(\omega)$, which is defined as the Fourier transform of the \emph{symmetrized current {\color{black} fluctuation} correlation function}, namely
\begin{equation}
\begin{split}
\label{eq:currentcf}
    S_{ab}(\omega)=\int_{-\infty}^{\infty}dt\;e^{i\omega t}\langle\delta\hat{J}_a(t)\delta\hat{J}_b(0)+\delta\hat{J}_b(0)\delta\hat{J}_a(t)\rangle,   
\end{split}
\end{equation}
where $\delta\hat{J}_a(t)=\hat{J}_a(t)-I_a$, with $I_a=\langle\hat{J}_a(t)\rangle$. Here, $S_{LL}$ ($S_{RR}$) is the noise power of the left (right) lead (auto-correlators), while $S_{LR}$ and $S_{RL}$ are called \emph{cross-noises} (cross-correlators). It is easy to check the parity property, $S_{ab}(-\omega)=S_{ba}(\omega)$, and the hermiticity of the noise matrix, $S^*_{ab}(\omega)=S_{ba}(\omega)$. 

{\color{black} The choice of the symmetrical noise over the non-symmetrical one is justified by the following reasons. Firstly, as we will see below, 
the symmetrized noise is easier to be derived within our theoretical formalism, secondly, it aligns closely with the classical interpretation of noise spectra, which are inherently symmetric. Moreover, symmetrized one is the most commonly measured quantity in standard noise detection experiments \cite{bib:blanter}. Most experimental setups, including spectrum analyzers and lock-in amplifiers, are designed to measure noise without distinguishing between emission (positive frequency) and absorption (negative frequency) processes \cite{bib:birk}. These devices average the noise power over both frequency directions, yielding a measurement that directly corresponds to the symmetrized noise spectrum. Specialized setups that measure asymmetrized noise exist (see for example \cite{bib:basset}) but require more complex instrumentation that can discern the directional dependence of quantum noise, which was beyond the scope of the present study.}

\section{KFT Formalism}
\label{sec:KFT}

We now present a the Keldysh field theory formalism, a powerful field-theoretical approach to describe quantum transport problems out-of-equilibrium. This is based on the Keldysh technique \cite{bib:keldysh}, that is a Green's functions technique which considers the evolution of a system along a closed real time contour $\mathcal{C}$, from $t=-\infty$ to $t=\infty$ and then going back. The main features of this formalism are presented in the self-contained book by Kamenev \cite{bib:kamenev}. The \emph{Keldysh action} for the two leads can be written in terms of fermionic coherent states, namely Grassmann variables $\bar\chi_{k\sigma a}$, $\bar\chi_{k\sigma a}$ which encode the operators $\hat{c}^\dagger_{k\sigma a}$, $\hat{c}_{k\sigma a}$, respectively, as
\begin{equation}
\begin{split}
    S_{leads}&=\sum_{a=L,R}\sum_k\sum_\sigma\int_\mathcal{C}dt\; \bar\chi_{k\sigma a}(t)(i\partial_t-\omega_{ka})\chi_{k\sigma a}(t)\\&=\sum_{a=L,R}\sum_k\sum_\sigma\int^\infty_{-\infty}dt\; \bigg[\bar\chi^+_{k\sigma a}(t)(i\partial_t-\omega_{ka})\chi^+_{k\sigma a}(t)+\\&-\bar\chi^-_{k\sigma a}(t)(i\partial_t-\omega_{ka})\chi^-_{k\sigma a}(t)\bigg]\\&=\sum_{a=L,R}\sum_k\sum_\sigma\int_\mathcal{-\infty}^\infty dtdt'\; \bar X^T_{k\sigma a}(t)\hat{g}^{-1}_{ka}(t-t')X_{k\sigma a}(t'),
\end{split}
\end{equation}
where the second line describes the fields along the Keldysh contour being expressed in terms of the $+,-$ branches, while in the third line a \emph{Keldysh rotation} has been applied, meaning $\chi^{1(2)}=(\chi^+\pm\chi^-)/\sqrt{2}$ and $\bar\chi^{1(2)}=(\bar\chi^+\mp\bar\chi^-)/\sqrt{2}$, where we introduce the compact vector in Keldysh space $X=\begin{pmatrix}
    \chi^1 & \chi^2
\end{pmatrix}^T$. Here, $\hat{g}^{-1}_{ka}$ is the inverse Green's function of the isolated leads in the $2\times 2$ Keldysh space, which reads in frequency space
\begin{equation}
    \hat{g}^{-1}_{ka}(\omega)=\begin{pmatrix}
        \omega-\omega_{ka}+i0^+ & 2i0^+F_a(\omega) \\ 0 & \omega-\omega_{ka}-i0^+
    \end{pmatrix},
\end{equation}
where infinitesimal $\pm i0^+$ are needed to express the physical coupling of the two branches at $t=-\infty$ given by the initial density matrix, while $F_a(\omega)$ is the distribution function for non-interacting fermions in the lead $a$, needed for the correct normalization of the \emph{Keldysh partition function}, $\mathcal{Z}=1$. Since we are assuming the leads to be at thermal equilibrium with temperatures $T_{a}$, the corresponding distribution functions are given by 
$F_{a}(\omega)=\tanh[\omega/(2T_{a})]\equiv 
{\color{black}1-2f_{a}(\omega)}$, being 
{\color{black}$f_a(\omega)=[e^{\omega/T_a}+1]^{-1}$}
the Fermi function. One can thus invert the matrix $\hat{g}_{ka}^{-1}$ to get the Green's function
\begin{equation}
   \hat{g}_{ka}= \begin{pmatrix} g^R_{ka} & g^K_{ka} \\ 0 & g^A_{ka}     
    \end{pmatrix}, 
\end{equation}
\begin{equation}
    g^{R(A)}_{ka}(\omega)=\frac{1}{\omega-\omega_{ka}\pm i0^+},
\end{equation}
\begin{equation}
\label{eq:FDT}
    g_{ka}^K(\omega)=F_a(\omega)[g_{ka}^R(\omega)-g_{ka}^A(\omega)],
\end{equation}
where the superscript $R,A,K$ stands for retarded, advanced, Keldysh Green's functions and, since the leads are at thermal equilibrium, these are related via the \emph{fluctuation-dissipation theorem} (FDT) in (\ref{eq:FDT}). Similarly the QD, described by the Grassmann fields $\bar\psi_{n\sigma}$, $\psi_{n\sigma}$ encoding the operators $\hat{d}^\dagger_{n\sigma}$, $\hat{d}_{n\sigma}$, will have the Keldysh action 
\begin{equation}
\begin{split} 
    S_{dot}&=\int_\mathcal{C} dt\; \bigg[\sum_n\sum_\sigma\Bar{\psi}_{n\sigma}(t)(i\partial_t-\varepsilon_n)\psi_{n\sigma}(t)-V(\Bar{\psi}_{n\sigma},\psi_{n\sigma})\bigg]\\&=\sum_n\sum_\sigma\int_{-\infty}^\infty dtdt'\;\bar\Psi^T_{n\sigma}(t)\hat{g}^{-1}_{n,0}(t-t')\Psi_{n\sigma}(t')+\\&-\int_{-\infty}^\infty dt\;V(\Bar{\psi}^1_{n\sigma},\psi^1_{n\sigma},\Bar{\psi}^2_{n\sigma},\psi^2_{n\sigma}),
\end{split}
\end{equation}
with the Keldysh vector $\Psi=\begin{pmatrix} \psi^1 & \psi^2\end{pmatrix}^T$ and the inverse Green's function matrix for the non-interacting QD
\begin{equation}
    \hat{g}^{-1}_{n,0}(\omega)=\begin{pmatrix}
     \omega-\varepsilon_n+i0^+  & 2i0^+F(\omega) \\ 0 & \omega-\varepsilon_n-i0^+
    \end{pmatrix},
\end{equation}
where $F(\omega)$ is an unknown distribution function for the non-interacting dot, which will be replaced by finite coupling to the bath later on. Finally, the tunneling action reads
\begin{equation}
\begin{split}
    S_{T}&=-\sum_{a=L,R}\sum_k\sum_n\sum_\sigma\int_\mathcal{C} dt\; \bigg[W_{ka,n}(t)\Bar{\chi}_{k\sigma a}(t)\psi_{n\sigma }(t) \\&+ W^*_{ka,n}(t)\Bar{\psi}_{n\sigma}(t)\chi_{k\sigma a}(t)\bigg],\\&=-\sum_{a=L,R}\sum_k\sum_n\sum_\sigma\int_{-\infty}^\infty dt\; \bigg[W_{ka,n}(t)\bar X^T_{k\sigma a}(t)\Psi_{n\sigma }(t) \\&+ W^*_{ka,n}(t)\Bar{\Psi}^T_{n\sigma}(t)X_{k\sigma a}(t)\bigg].
\end{split}
\end{equation} 

By construction of the Keldysh field theory, the Keldysh partition function $\mathcal{Z}$ is normalized by construction
\begin{equation}
    \mathcal{Z}=\int D[\bar X,X]D[\bar \Psi,\Psi]e^{iS}\equiv 1 
\end{equation}
{\color{black} where $S=S_{dot}+S_{leads}+S_{T}$}. 
To make the entire theory meaningful one should introduce auxiliary source fields, which enable one to compute various observable quantities. In our case we introduce two source fields $A_a$, interacting with the currents
\begin{equation}
\label{eq:Jfields}
    J_a=ie\sum_{k}\sum_n\sum_\sigma\bigg[W_{ka,n}\bar\chi_{k\sigma a}\psi_{n\sigma}-W^*_{ka,n}\bar\psi_{n\sigma}\chi_{ka\sigma}\bigg],
\end{equation}
encoding the operators $\hat{J}_a$ in (\ref{eq:Jtransport}), with an action
\begin{equation}
\begin{split}
    S_A&=-\sum_{a=L,R}\int_{\mathcal{C}}dt\; A_{a}(t)J_a(t)\\&=-\sum_{a=L,R}\int^{\infty}_{-\infty}dt\; \bigg[A^+_{a}(t)J^+_a(t)-A^-_{a}(t)J^-_a(t)\bigg]\\&=-2\sum_{a=L,R}\int^{\infty}_{-\infty}dt\; \bigg[A^{cl}_{a}(t)J^q_a(t)-A^q_{a}(t)J^{cl}_a(t)\bigg],
\end{split}
\end{equation}
where we introduced the classical and quantum components of the fields $J^{cl(q)}=(J^+\pm J^-)/2$ and $A^{cl(q)}=(A^+\pm A^-)/2$. We see that the physical currents (symmetrized over the two branches of the contour) are $J_a^{cl}$, and are coupled to the quantum component of the source fields $A_a^q$ On the other hand, the quantum components of the currents $J^q_a$ are coupled to the classical source components, $A^{cl}_a$, which are nothing but external physical potentials, the same on the two branches. Since we want to generate the physical currents and we are not interested in any external potential (the external voltage $V$ has already been included in the tunneling amplitudes), we can set $A_a^{cl}=0$ and thus we can explicitly write
\begin{equation}
\begin{split}
    &S_A=-ie\sum_{a=L,R}\sum_k\sum_n\sum_\sigma\int_{-\infty}^\infty dt\; A^q_a(t)\times\\&\bigg[W_{ka,n}(t)\Bar{X}^T_{k\sigma a}(t)\hat{\tau}_x\Psi_{n\sigma}(t)-W^*_{ka,n}(t)\Bar{\Psi}^T_{n\sigma}(t)\hat{\tau}_xX_{k\sigma a}(t)\bigg],
\end{split}
\end{equation}
where $\hat{\tau}_x$ is the first Pauli matrix in Keldysh space, namely
\begin{equation}
    \hat{\tau}_x=\begin{pmatrix}
        0 & 1 \\ 1 & 0
    \end{pmatrix}.
\end{equation}
This way, the expectation values of the currents (\ref{eq:Jfields}) are given by {\cite{bib:kamenev}}
\begin{equation}
\label{eq:I_a}
    I_a=\langle J^{cl}_a(t)\rangle=\frac{i}{2}\frac{\delta}{\delta A_a^q(t)}\mathcal{Z}[A_a^q]\bigg|_{A_a^q=0},
\end{equation}
where $\mathcal{Z}[A^q_a]$ is the \emph{Keldysh generating functional}, given by 
\begin{equation}
    \mathcal{Z}[A^q_a]=\int D[\bar{\chi},\chi]D[\bar{\psi},\psi]e^{iS+iS_A},
\end{equation}
and if the quantum sources are set to zero, namely $A_a^q=0$, one recovers normalized Keldysh partition function, $\mathcal{Z}=\mathcal{Z}[0]=1$. Similarly, we can easily derive the second cumulant of the current distribution from the Keldysh action, by taking the second functional derivative of the logarithm of the Keldysh partition function with respect to the quantum source $A_a^q$. Namely
\begin{equation}
\label{eq:noiseA}
    S_{ab}(\omega)= \frac{i^2}{2}\frac{\delta^2}{\delta A_a^q(\omega)\delta A_b^q(-\omega)}\ln\mathcal{Z}[A_a^q]\bigg|_{A_a^q=0},
\end{equation}
which automatically gives the properly symmetrized noise power (\ref{eq:currentcf}) {\cite{bib:kamenev}}.

We can now use the fact that the leads are non-interacting to perform the Gaussian integrations over the $\bar{X}_{k\sigma a}$, {$X_{k\sigma a}$} variables and to find the effective action
\begin{equation}
\begin{split}
\label{eq:s_eff_dot}
    S_{eff}&=S_{dot}-\sum_{a}\sum_{n,m}\sum_\sigma\int_{-\infty}^\infty dt\;dt'\;\Bar{\Psi}^T_{n\sigma}(t)[1-ieA^q_a(t)\hat{\tau}_x]\\&\times\hat{\Sigma}_{a,nm}(t,t')[1+ieA^q_a(t')\hat{\tau}_x]\Psi_{m\sigma}(t'),
\end{split}
\end{equation}
where
\begin{equation}
    \hat{\Sigma}_{a,nm}(t,t')=\sum_k W_{ka,n}^*(t)\hat{g}_{ka}(t-t')W_{ka,m}(t'),
\end{equation}
is a self-energy contribution produced by the coupling of the QD with the leads. Notice that, in frequency space, the effects of the time dependent phase $e^{i\mu_a (t-t')}$ consist simply in {\color{black} an energy shift in the Green's function of the uncoupled leads}, $\hat{g}_{ka}(\omega)\to\hat{g}_{ka}(\omega-eV)$. Since the leads are assumed to be in local equilibrium, the FDT dictates that
\begin{equation}
   \Sigma^{K}_{a,nm}(\omega)=F_a(\omega-\mu_a)[\Sigma^{R}_{a,nm}(\omega)-\Sigma^{A}_{a,nm}(\omega)]. 
\end{equation}
We therefore differentiate the effective action over the quantum sources, as in (\ref{eq:I_a}), to get a stationary average current (we restore $\hbar$ in the expression) 
\begin{equation}
\begin{split}
\label{eq:I^A_a}
    I_a&=\frac{e}{\hbar}\sum_{n,m}\int_{-\infty}^\infty\frac{d\omega}{2\pi}\mbox{tr}\big\{[\hat{\Sigma}_{a,nm}(\omega),\hat{\tau}_x]\hat{G}_{mn}(\omega)\big\}\\&=\frac{ie}{h}\sum_{n,m}\int_{-\infty}^{\infty}d\omega\; \Gamma^a_{nm}(\omega-\mu_a)\bigg[G_{mn}^K(\omega)+\\&-F_a(\omega-\mu_a)\big(G^R_{mn}(\omega)-G^A_{mn}(\omega)\big)\bigg],
    \end{split}
\end{equation}
where $\hat{G}_{nm}=-i\langle\Psi_n\Bar{\Psi}^T_m\rangle_{S_{eff}[A^q_a=0]}$ are the components of the exact Green's function of the coupled QD, 
and we introduced the couplings
\begin{equation}
    \Gamma^a_{nm}(\omega)=2\pi\sum_k W^*_{ka,n}\delta(\omega-\omega_{ka})W_{ka,m},
\end{equation}
which provide us energy scales characterizing the strength of the coupling between the QD and the leads. In particular it is easy to check that, if the dot and the two leads are in equilibrium, then $I_L=I_R=0$. In a non-equilibrium stationary state $I_L=-I_R$. This leads to the expression for the DC current through the dot, $I=(I_L-I_R)/2$,
\begin{equation}
\begin{split}
\label{eq:meir-wingreen}
    I&=\frac{ie}{2h}\sum_{n,m}\int_{-\infty}^{\infty}d\omega\; \bigg[\big(\Gamma^L_{nm}(\omega-eV)-\Gamma^R_{nm}(\omega)\big)G_{mn}^K(\omega)+\\&-\big(F_L(\omega-eV)\Gamma^L_{nm}(\omega-eV)-F_R(\omega)\Gamma^R_{nm}(\omega)\big)\times\\&\big(G^R_{mn}(\omega)-G^A_{mn}(\omega)\big)\bigg].
    \end{split}
\end{equation}
This result, known as \emph{Meir-Wingreen formula} and originally derived using the Keldysh formalism in its operatorial version \cite{bib:meir,bib:wingreen}, expresses the current through the coupling matrices $\Gamma^a_{nm}(\omega)$ and the exact Green's functions of a quantum dot. The latter refer not to the isolated dot, but rather to the dot coupled to the leads. Notice that this expression is very general, provided that we are dealing with non-interacting leads. It is also interesting to notice that, if the two leads are at different temperature, the current is non-zero even in the case $V=0$. This means that the Meir-Wingreen formula derived above also includes the thermoelectric effects or Seebeck effects, namely the creation of a current due to a temperature gradient.

One can make a step forward and also compute the noise power matrix $S_{ab}(\omega,V)$ by using (\ref{eq:noiseA}). In this paper, we assume a quadratic QD action, namely
\begin{equation}
\label{eq:quadraticdot}
    S_{dot}=\sum_{nm}\sum_\sigma\int_{-\infty}^\infty dt\; \Bar{\Psi}^T_{n\sigma}(t)\hat{g}^{-1}_{nm}(t-t')\Psi_{m\sigma}(t'),
\end{equation}
\begin{equation}
    \hat{g}^{-1}_{nm}(t,t')=\bigg[\delta_{nm}\hat{g}^{-1}_{n,0}(t,t')-\hat{\Sigma}_{dot,nm}(t,t')\bigg]
\end{equation}
where $\hat{g}_{nm}$ includes also the possible effects of the interactions inside the dot, described by a generic self-energy matrix $\hat{\Sigma}_{dot,nm}$, which has the same structure of $\hat{g}_{n,0}$ (also known as \emph{causality structure} \cite{bib:kamenev}). This assumption is compatible with interacting systems, provided that the QD two-particle Green's function can be decoupled into a product of two QD one-particle Green's function. Speaking in terms of Feynman diagrams, this statement equivalent to consider the vertex corrections as negligible. Notice that the total self-energy for the coupled QD is thus $\hat{\Sigma}_{nm}=\hat{\Sigma}_{dot,nm}+\hat{\Sigma}_{L,nm}+\hat{\Sigma}_{R,nm}$ in this problem. Within this condition, we can integrate further the effective action of the dot (\ref{eq:s_eff_dot}), obtaining
\begin{equation}
\begin{split}
    \ln\mathcal{Z}[A^q]&=\Tr \ln\bigg[\hat{1}-\hat{G}\bigg(ie\hat{\Sigma}_{a}\hat{A}_a^q-ie\hat{A}_a^q\hat{\Sigma}_{a}+e^2\hat{A}_a^q\hat{\Sigma}_{a}\hat{A}_a^q\bigg)\bigg]\\ &\simeq  -\Tr \bigg[ie\hat{G}\bigg(\hat{\Sigma}_{a}\hat{A}_a^q-\hat{A}_a^q\hat{\Sigma}_{a}\bigg)+\\&+\frac{e^2}{2}\hat{G}\bigg(2\hat{A}_{a}^q\hat{\Sigma}_{a}\hat{A}_a^q+2\hat{A}_a^q\hat{\Sigma}_{a}\hat{G}\hat{\Sigma}_{b}\hat{A}_b^q +\\&-\hat{\Sigma}_a\hat{A}_a^q\hat{G}\hat{\Sigma}_b\hat{A}_b^q-\hat{A}_a^q\hat{\Sigma}_a\hat{G}\hat{A}_b^q\hat{\Sigma}_b\bigg)+O([A_a^q]^3)\bigg],
\end{split}
\end{equation}
where $\hat{A}_a^q=A_a^q\hat{\tau}_x$, and a summation of the repeated lead index $a$ is left implicit. Notice that here the trace includes also the sum over the dot's levels $n,m$ and so on.  Expanding this expression at first order in $\hat{A}_a^q$ allows us to recover the current (\ref{eq:I^A_a}), while expanding at second order gives us the full expression for the noise. In particular, by differentiating twice over the quantum source, we get the noise power
\begin{equation}
\begin{split}
\label{eq:S_ab}
    S_{ab}(\omega)&=\frac{e^2}{h} \int_{-\infty}^{\infty}d\epsilon\;\mbox{tr}\bigg[\delta_{ab}\hat{G}(\epsilon_+)\hat{\tau}_x\hat{\Sigma}_a(\epsilon_-)\hat{\tau}_x+\\&+\delta_{ab}\hat{G}(\epsilon_-)\hat{\tau}_x\hat{\Sigma}_a(\epsilon_+)\hat{\tau}_x+\\&+\hat{G}(\epsilon_+)\hat{\tau}_x\hat{\Sigma}_a(\epsilon_-)\hat{G}(\epsilon_-)\hat{\Sigma}_b(\epsilon_-)\hat{\tau}_x+\\&+\hat{G}(\epsilon_-)\hat{\tau}_x\hat{\Sigma}_b(\epsilon_+)\hat{G}(\epsilon_+)\hat{\Sigma}_a(\epsilon_+)\hat{\tau}_x+\\&-\hat{G}(\epsilon_+)\hat{\Sigma}_a(\epsilon_+)\hat{\tau}_x\hat{G}(\epsilon_-)\hat{\Sigma}_b(\epsilon_-)\hat{\tau}_x+\\&-\hat{\Sigma}_b(\epsilon_+)\hat{G}(\epsilon_+)\hat{\tau}_x\hat{\Sigma}_a(\epsilon_-)\hat{G}(\epsilon_-)\hat{\tau}_x\bigg],
\end{split}
\end{equation}
where $\epsilon_\pm=\epsilon\pm \omega/2$, and we see that the parity property $S_{ab}(-\omega)=S_{ba}(\omega)$ is clearly satisfied. The exact Green's function of the coupled QD, $\hat{G}_{nm}$,  can be obtained via the \emph{Keldysh-Dyson equation}, which reads in frequency space
\begin{equation}
    \label{eq:Keldysh-Dyson}
\hat{G}_{nm}(\omega)=\hat{g}_{nm}(\omega)+\sum_{a=L,R}\sum_{ij}\hat{g}_{ni}(\omega)\hat{\Sigma}_{a,ij}(\omega)\hat{G}_{jm}(\omega).
\end{equation} Because of the trace also over the dot indices and the double-frequency dependence in the expressions, the evaluation of the noise power can be quite cumbersome in general. In the following, we will analyze the full noise only for the case of a single-level QD, while we will just focus on the current in the case of a two-level QD.

\section{Single-level QD}
\label{bib:single}

We consider the simple case of a non-interacting dot with a single discrete level $\varepsilon_0$ coupled to two metallic leads, with $k$-independent tunneling matrix elements $W_{L}$ and $W_{R}$. By using the \emph{Keldysh-Dyson eqution} (\ref{eq:Keldysh-Dyson}), the retarded (advanced) Green's function of the coupled dot can be written as
\begin{equation}
\label{eq:G^R_00}
    G^{R(A)}_{00}(\omega)=\frac{1}{\omega-\varepsilon_0-|W_L|^2g^{R(A)}_{L}(\omega-eV)-|W_R|^2g^{R(A)}_{R}(\omega)},
\end{equation}
where $g^{R(A)}_a(\omega)=\sum_k g^{R(A)}_{ka}(\omega)$, while the Keldysh Green's function is given by
\begin{equation}
\begin{split}
\label{eq:G^K_00}
    G^{K}_{00}(\omega)&=-2\pi i\bigg[F_L(\omega-eV)|W_{L}|^2\nu_L(\omega-eV)+\\&+F_R(\omega)|W_{R}|^2\nu_R(\omega)\bigg]|G^R_{00}(\omega)|^2,
\end{split}
\end{equation}
with \emph{density of states} (DOS) $\nu_a(\omega)=\sum_k\delta(\omega-\omega_{ka})$. 

\subsection{Current} 

By substituting $G^{R(A)}_{00}$ and $G^K_{00}$ expressions in the Meir-Wingreen formula (\ref{eq:meir-wingreen}), one gets an expression for the average current in the form of a generalized \emph{Landauer formula} \cite{bib:landauer}, namely
\begin{equation}
\label{eq:genland2}
    I=\frac{2e}{h}\int_{-\infty}^{\infty}d\omega\; \big[f_L(\omega-eV)-f_R(\omega)\big]\mathcal{T}(\omega,V),
\end{equation}
where the \emph{transmission coefficient} $\mathcal{T}$ is given by
\begin{equation}
\label{eq:transmcoeff}
    \mathcal{T}(\omega,V)=\frac{4\pi^2|W_L|^2|W_R|^2\nu_L(\omega-eV)\nu_R(\omega)}{|\omega-\varepsilon_0-|W_L|^2g^{R}_{L}(\omega-eV)-|W_R|^2g^{R}_{R}(\omega)|^2}.
\end{equation} 

\begin{figure}[t!]
  \includegraphics[width=\linewidth]{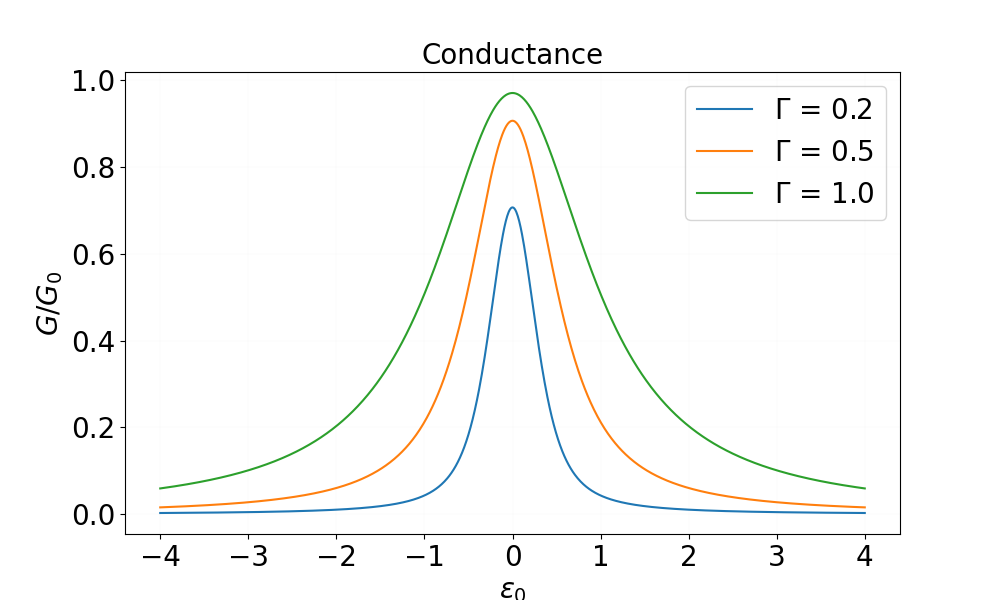}
  \caption{Linear conductance $G$ in units of the conductance quantum $G_0=2e^2/h$ as a function of the dot's energy $\varepsilon_0$ at fixed temperature ($\beta=10$) for different values of the coupling $\Gamma{\color{black}=\Gamma^L=\Gamma^R}$. 
}
  \label{fig:1}
  \includegraphics[width=\linewidth]{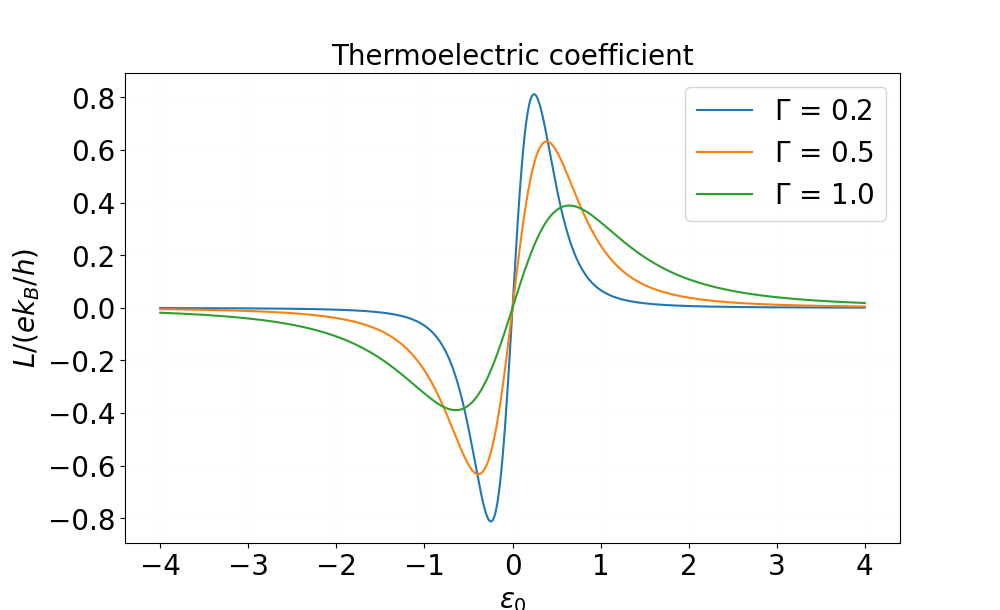}
  \caption{Thermoelectric coefficient $L$ in units of {\color{black} $ek_B/h$} as a function of the dot's energy $\varepsilon_0$ at fixed temperature ($\beta=10$) for different values of the coupling $\Gamma{=\Gamma^L=\Gamma^R}$.}
  \label{fig:2}
\end{figure}

In the \emph{linear response regime}, namely $V$ and $\Delta T$ small (where $\Delta T=T_L-T_R>0$ is the temperature difference between the leads), the Fermi function can be expanded as
\begin{equation}
    f_L(\omega-eV)\simeq f_R(\omega)-\frac{df_L}{d\omega}(\omega)\bigg[eV+\frac{\omega}{T_L}\Delta T\bigg],
\end{equation}
and the current reads $I=GV+L\Delta T$, where $G$ is the \emph{conductance} and $L$ is the \emph{thermoelectric coefficient}. In this regime, for scales near the Fermi energy, one can also use the \emph{wide-band approximation} \cite{bib:wingreen}, which consists in neglecting the real part (energy level shift) of the Green's functions for the uncoupled leads and considering a constant DOS evaluated at the Fermi energy $\nu_a$, namely $g^{R}_{a}=-i\pi\nu_a$. In this way, we find 
\begin{equation}
\label{eq:GNQDN}
    G=-\frac{2e^2}{h}\Gamma^L\Gamma^R\int_{-\infty}^{\infty}d\omega\;\frac{\frac{df_L}{d\omega}(\omega)}{(\omega-\varepsilon_0)^2+(\Gamma^L/2+\Gamma^R/2)^2},
\end{equation}
where the constant couplings are $\Gamma^a=2\pi|W_a|^2\nu_a$. In particular, at zero temperature, the integral can be easily performed, obtaining a standard Landauer formula \cite{bib:landauer}
\begin{equation}
    G=\frac{2e^2}{h}\frac{\Gamma^L\Gamma^R}{\varepsilon_0^2+(\Gamma^L/2+\Gamma^R/2)^2}.
\end{equation}
The two equations above predicts that the conductance in the dot is maximum when $\varepsilon_0=0$ (i.e. at the Fermi energy), which is the condition for \emph{resonant tunneling}. For a \emph{symmetric junction}, $\Gamma^R=\Gamma^L$, this maximum value at zero temperature equals the \emph{conductance quantum} $G_0=2e^2/h$, irrespective of the strength of the coupling to the electrodes. A plot of the linear conductance $G$ as a function of the dot's energy $\varepsilon_0$ {at fixed $\beta=(1/k_B T_L)$} is presented in Fig. \ref{fig:1} for three different values of the symmetric coupling $\Gamma{=\Gamma^L=\Gamma^R}$. We clearly see that the conductance maximum at $\varepsilon_0=0$ is increased for higher values of the coupling $\Gamma$. 

On the other hand, one gets a thermoelectric coefficient
\begin{equation}
\label{thermalcoeff}
    L=-\frac{2ek_B}{h}\Gamma^L\Gamma^R\int_{-\infty}^{\infty}d\omega\;\frac{\beta\omega\,\frac{df_L}{d\omega}(\omega)}{(\omega-\varepsilon_0)^2+(\Gamma^L/2+\Gamma^R/2)^2},
\end{equation}
which is always zero at the Fermi energy $\varepsilon_0=0$. This result is compatible with the fact that \emph{particle-hole symmetry} in a system can strongly influence thermoelectric effects, with an ideal net cancellation of thermoelectric currents, as in this case \cite{bib:marchegiani}. This cancellation occurs because for every electron that contributes to the current in one direction due to the temperature gradient, there would be an exactly compensating hole contribution in the opposite direction. Moreover, one can easily check that the thermoelectric coefficient $L$ is an odd function in the dot's energy $\varepsilon_0$. This means that one can control the direction of the thermoelectric current by moving the level of the dot below or above the Fermi energy, as already shown both experimentally \cite{bib:9} and theoretically by other techniques and setups \cite{bib:therm1, bib:therm2,bib:therm3, bib:therm4,bib:therm5, bib:therm6}.
The plot of the thermoelectric coefficient $L$ as a function of the dot's energy $\varepsilon_0$ is reported in Fig. \ref{fig:2}. We clearly see that the thermoelectric coefficient is always zero at the Fermi energy,  as discussed above, and there exist two resonant peaks, for each value of the coupling $\Gamma$, for which the modulus of the thermoelectric coefficient is maximal.

{\color{black} Let us consider, for simplicity, the symmetric case $\Gamma^L=\Gamma^R=\Gamma$. We first notice that Eqs.~(\ref{eq:GNQDN}), (\ref{thermalcoeff}) are invariant under $\beta\rightarrow 1$ and rescaling
\begin{equation}
    \Gamma\rightarrow \beta\Gamma,\qquad \varepsilon_0\rightarrow \beta \varepsilon_0
\end{equation}
namely we can absorb the temperature dependence in the coupling and energy parameters, $L(\beta, \Gamma, \varepsilon_0)= L(\beta \Gamma, \beta\varepsilon_0)$.\\
In the weak coupling or large temperature regime, namely for $\beta\Gamma\ll 1$, we can replace the Cauchy-Lorentz distribution with a regularized delta function, $\frac{\Gamma}{\pi}\frac{1}{(\omega-\varepsilon_0)^2+\Gamma^2}\rightarrow \delta_\Gamma(\omega-\varepsilon_0)$, so that, from Eq.~(\ref{thermalcoeff}), we get 

\begin{equation}
L\simeq \frac{ek_B}{\hbar}\, \beta\Gamma  \frac{\beta \varepsilon_0\,e^{\beta\varepsilon_0}}
{(1+e^{\beta\varepsilon_0})^2},
\end{equation}
As a result, in the weak coupling limit, the resonant peaks are placed at $\varepsilon_0=\bar\varepsilon_0$ such that 
\begin{equation}
\tanh\left(\frac{\beta\bar\varepsilon_0}{2}\right)=\frac{1}{\beta\bar\varepsilon_0},
\end{equation}
whose solution are
\begin{equation}
\beta\bar\varepsilon_0\approx \pm 1.54.
\end{equation}
At weak coupling, the thermoelectric coefficient reaches its maximum absolute value when the quantum dot energy level is tuned at $\varepsilon_0\approx \pm 1.54\, k_B T$, where $T=T_L\approx T_R$ since we are in the linear response regime (small $\Delta T$).
The maximum value of the thermoelectric coefficient in the weak coupling regime is, therefore, $L_{max}\approx 0.22\frac{ek_B}{\hbar}\, \beta\Gamma$, namely it becomes smaller and smaller going deeply in the weak coupling limit.

On the other hand, as shown in Fig. \ref{fig:2}, in the strong coupling regime, $\beta \Gamma\gg 1$, the thermoelectric coefficient is again suppressed because the integrand in Eq. (\ref{thermalcoeff}) becomes more and more antisymmetric (namely, it becomes less sensitive to the dot energy level). 

We have shown that in both weak and strong coupling regimes the thermoelectric coefficient is suppressed, therefore, we expect that in the intermediate regime, $\beta\Gamma\sim 1$,  it reaches its maximum value. 
After rescaling one can easily show numerically that
\begin{equation}
L=\frac{2ek_B}{h}\,\int^{\infty}_{-\infty} dx \frac{x\,e^x}{(1+e^x)^2}\frac{(\beta\Gamma)^2}{(x-\beta\varepsilon_0)^2+(\beta\Gamma)^2},
\end{equation}
reaches its maximum absolute values for
\begin{equation}
    \beta\Gamma \approx 2, \qquad 
\beta\varepsilon_0\approx \pm 2.5.
\end{equation}
The blue curve reported in Fig. \ref{fig:2} is obtained for $\beta\Gamma=2$. 
We have therefore that, for those values of the coupling and dot energy level, one reaches the maximum value for $L$ which is
\begin{equation}
\label{Lm}
L_{max} \approx 0.8 \frac{ek_B}{h}.
\end{equation}
We show, therefore, that the thermoelectric transport is made maximally efficient tuning the parameters 
so that $\varepsilon_0\approx \pm 1.2\Gamma$ and in a temperature regime $k_BT\approx \Gamma/2$. 
Moreover we find that the thermoelectric coefficient has a universal threshold given by  Eq. (\ref{Lm}).
}

\subsection{Noise}

By performing the trace over the Keldysh space in (\ref{eq:S_ab}), we get the expression of the noise matrix for a single-level QD 
\begin{equation}
\begin{split}
\label{eq:whitenoise}
    &S_{ab}(\omega)=\frac{e^2}{h} \int_{-\infty}^{\infty}d\epsilon\;\bigg\{i\delta_{ab}\Gamma^a_-\bigg[2i\Im G_{+}^R-G_{+}^KF_{a-}\bigg]\\&+i\delta_{ab}\Gamma^a_+\bigg[2i\Im G_{-}^R-G_{-}^KF_{a+}\bigg]\\&+ G_+^KG_-^K\bigg[\Sigma^R_{a-}\Sigma^A_{b-}+\Sigma^R_{a+}\Sigma^A_{b+}-\Sigma^R_{a-}\Sigma^R_{b+}-\Sigma^A_{a+}\Sigma^A_{b-}\bigg]+
    \\&+ G_+^RG_-^A\bigg[\Sigma^A_{a-}\Sigma^A_{b-}+\Sigma^R_{a+}\Sigma^R_{b+}-\Sigma^R_{a+}\Sigma^A_{b-}-\Sigma^A_{a-}\Sigma^R_{b+}\bigg]+\\&+G_-^RG_+^A\bigg[\Sigma^A_{a+}\Sigma^A_{b+}+\Sigma^R_{a-}\Sigma^R_{b-}-\Sigma^A_{a+}\Sigma^R_{b-}-\Sigma^R_{a-}\Sigma^A_{b+}\bigg]+\\&-iG^K_+\bigg[G_-^R\bigg(\Sigma^R_{a-}-\Sigma^A_{a+}\bigg)F_{b-}\Gamma^b_-+G^A_-\bigg(\Sigma^A_{b-}-\Sigma^R_{b+}\bigg)F_{a-}\Gamma^a_-\bigg]\\&-iG^K_-\bigg[G^A_+\bigg(\Sigma^A_{a+}-\Sigma^R_{a-}\bigg)F_{b+}\Gamma^b_++G^R_+\bigg(\Sigma^R_{b+}-\Sigma^A_{b-}\bigg)F_{a+}\Gamma^a_+\bigg]\\&+F_{a+}F_{b-}\Gamma^a_+\Gamma^b_-G^R_+G^R_-+F_{a-}F_{b+}\Gamma^a_-\Gamma^b_+G^A_+G^A_-\bigg\},
\end{split}
\end{equation}
where we neglected the explicit argument of the functions inside the integral, leaving only the notation $\pm$ to denote the argument $\epsilon_\pm=\epsilon\pm \omega/2$. From this expression one can clearly check that the hermiticity of the matrix is satisfied, $S^*_{ab}(\omega)=S_{ba}(\omega)$, together with the parity property $S_{ab}(-\omega)=S_{ba}(\omega)$. As one can see, the expression is quite long and cumbersome, but one can express the solution in a relatively simple form by working in the wide-band limit for a symmetric junction, $\Gamma^R=\Gamma^L=\Gamma$. By using the explicit form of $G^R_{00}$ and $G^K_{00}$, obtained in (\ref{eq:G^R_00}) and (\ref{eq:G^K_00}) respectively, and the constant self-energies $\Sigma^R_L=\Sigma^R_R=-i\Gamma/2$, one can easily get the following noise matrix elements 
\begin{equation}
\begin{split}
   &S_{LL}(\omega,V)=\frac{e^2}{h} \int_{-\infty}^{\infty}d\epsilon\;\bigg[\mathcal{T}(\epsilon_+)\big(1-\mathcal{T}(\epsilon_-)\big)B_{RL}(\epsilon,\omega)+\\&+\mathcal{T}(\epsilon_-)\big(1-\mathcal{T}(\epsilon_+)\big)B_{LR}(\epsilon,\omega)+\mathcal{T}(\epsilon_+)\mathcal{T}(\epsilon_-)B_{RR}(\epsilon,\omega)+\\&+\big(\mathcal{T}(\epsilon_+)\mathcal{T}(\epsilon_-)+|t(\epsilon_+)-t(\epsilon_-)|^2\big)B_{LL}(\epsilon,\omega),
\end{split}
\end{equation}
\begin{equation}
\begin{split}
   &S_{RR}(\omega,V)=\frac{e^2}{h} \int_{-\infty}^{\infty}d\epsilon\;\bigg[\mathcal{T}(\epsilon_+)\big(1-\mathcal{T}(\epsilon_-)\big)B_{LR}(\epsilon,\omega)+\\&+\mathcal{T}(\epsilon_-)\big(1-\mathcal{T}(\epsilon_+)\big)B_{RL}(\epsilon,\omega)+\mathcal{T}(\epsilon_+)\mathcal{T}(\epsilon_-)B_{LL}(\epsilon,\omega)+\\&+\big(\mathcal{T}(\epsilon_+)\mathcal{T}(\epsilon_-)+|t(\epsilon_+)-t(\epsilon_-)|^2\big)B_{RR}(\epsilon,\omega)\bigg],
\end{split}
\end{equation}
\begin{equation}
\begin{split}
    &S_{LR}(\omega,V)=\frac{e^2}{h} \int_{-\infty}^{\infty}d\omega\;\bigg[t(\epsilon_+)t^*(\epsilon_-)\big[\big(1-t(\epsilon_-)\big)\times\\&(1-t^*(\epsilon_+))-1\big]B_{LL}(\epsilon,\omega)+t^*(\epsilon_+)t(\epsilon_-)\big[\big(1-t^*(\epsilon_-)\big)\times\\& (1-t(\epsilon_+))-1\big]B_{RR}(\epsilon,\omega)+t(\epsilon_+)t(\epsilon_-)\big[\big(1-t^*(\epsilon_+)\big)\times\\&\big(1-t^*(\epsilon_-)\big)\big]B_{LR}(\epsilon,\omega)++t^*(\epsilon_+)t^*(\epsilon_-)\big[\big(1-t(\epsilon_+)\big)\times\\&\big(1-t(\epsilon_-)\big)\big]B_{RL}(\epsilon,\omega)\bigg],
\end{split}
\end{equation}
\begin{equation}
\begin{split}
    &S_{RL}(\omega,V)=\frac{e^2}{h} \int_{-\infty}^{\infty}d\omega\;\bigg[t(\epsilon_+)t^*(\epsilon_-)\big[\big(1-t(\epsilon_-)\big)\times\\&(1-t^*(\epsilon_+))-1\big]B_{RR}(\epsilon,\omega)+t^*(\epsilon_+)t(\epsilon_-)\big[\big(1-t^*(\epsilon_-)\big)\times\\& (1-t(\epsilon_+))-1\big]B_{LL}(\epsilon,\omega)+t(\epsilon_+)t(\epsilon_-)\big[\big(1-t^*(\epsilon_+)\big)\times\\&\big(1-t^*(\epsilon_-)\big)\big]B_{RL}(\epsilon,\omega)++t^*(\epsilon_+)t^*(\epsilon_-)\big[\big(1-t(\epsilon_+)\big)\times\\&\big(1-t(\epsilon_-)\big)\big]B_{LR}(\epsilon,\omega)\bigg],
\end{split}
\end{equation}
where we introduced the four statistical factors $B_{ab}$, defined as $B_{ab}(\epsilon,\omega)=1-F_a(\epsilon_+-\mu_a)F_b(\epsilon_--\mu_b)$. Here $\mathcal{T}$ is the transmission coefficient introduced in (\ref{eq:genland2}), which in the wide-band limit for a symmetric junctions reads
\begin{equation}
    \mathcal{T}(\omega)=\Gamma^2  {|G^R_{00}(\omega)|^2}=\frac{\Gamma^2}{(\omega-\varepsilon_0)^2+\Gamma^2},
\end{equation} 
and we also introduced the \emph{transmission amplitude} $t(\omega)$, defined as
\begin{equation}
    t(\omega)=i\Gamma G^R_{00}(\omega)=\frac{i\Gamma}{\omega-\varepsilon_0+i\Gamma},
\end{equation}
such that $\mathcal{T}(\omega)=t(\omega)t^*(\omega)=\Re t(\omega)$. We notice that when both finite-frequency and non-equilibrium are considered, all the  noise matrix elements differ from each other. These results are perfectly compatible with the ones obtained by Zamoum et al. \cite{bib:zamoum} within the operatorial version of the Keldysh formalism. Moreover, we can compute the sum of the four noises
\begin{equation}
    \sum_{ab=L,R}S_{ab}(\omega,V)=\frac{e^2}{h}\int_{-\infty}^\infty d\epsilon\; |t(\epsilon_+)-t(\epsilon_-)|^2\sum_{ab=L,R}B_{ab}(\epsilon,\omega),
\end{equation}
which is always a real quantity, and is related to the \emph{charge
fluctuations} in the QD via the relation
\begin{equation}
\label{eq:chargefluc}
    \sum_{ab=L,R}S_{ab}(\omega)=\omega^2S_Q(\omega),
\end{equation}
as shown in Appendix within our field-theoretical formalism.

Let us consider two limiting cases. If the two leads are at the same temperature $T$, one can obtain the \emph{thermal noise} or \emph{Johnson-Nyquist noise}, corresponding to equilibrium current fluctuations ($V\to0$), as
\begin{equation}
\begin{split}
    &S_{LL}(\omega,0)=S_{RR}(\omega,0)=\frac{2e^2}{h}\Gamma^2\int_{-\infty}^{\infty}d\epsilon\;\\&\bigg[\frac{\omega^2+\omega(\epsilon-\varepsilon_0)+(\epsilon-\varepsilon_0)^2+\Gamma^2}{[(\epsilon+\omega-\varepsilon_0)^2+\Gamma^2][(\epsilon-\varepsilon_0)^2+\Gamma^2]}\bigg]\big(1-F(\epsilon+\omega)F(\epsilon)\big),    
    \end{split}
\end{equation}
\begin{equation}
\begin{split}
    &S_{LR}(\omega,0)=S_{RL}(\omega,0)=-\frac{2e^2}{h}\Gamma^2\int_{-\infty}^{\infty}d\epsilon\;\\&\bigg[\frac{\omega(\epsilon-\varepsilon_0)+(\epsilon-\varepsilon_0)^2+\Gamma^2}{[(\epsilon+\omega-\varepsilon_0)^2+\Gamma^2][(\epsilon-\varepsilon_0)^2+\Gamma^2]}\bigg]\big(1-F(\epsilon+\omega)F(\epsilon)\big),    
    \end{split}
\end{equation}
which has a simple form in the zero frequency limit, namely
\begin{equation}
    S_{LL}(0,0)=-S_{LR}(0,0)=4Gk_BT
\end{equation}
where we used the expression of the linear conductance (\ref{eq:GNQDN}), and the identity $1-F^2(\omega)=-4/\beta \frac{df}{d\omega}(\omega)$. The result above is exactly the classical version of the FDT \cite{bib:callen}. Similarly, one can get the non-equilibrium noise, or \emph{shot noise} ($T=0$), by noticing that the effects of the statistical factors $B_{ab}$ on a generic test function $g$ at zero temperature are given by the following relation
\begin{equation}
    \int_{-\infty}^{\infty}d\epsilon\; B_{ab}(\epsilon,\omega)g(\epsilon)=2\mbox{sgn}(\omega-\mu_a+\mu_b)\int_{\mu_a-\frac{\omega}{2}}^{\mu_b+\frac{\omega}{2}}d\epsilon\;g(\epsilon).
\end{equation}
In particular, the shot noise has the following analytical expression in the zero-frequency limit
\begin{equation}
\begin{split}
    S_{LL}(0,V)&=S_{RR}(0,V)=\frac{4e^2}{h}\int_{0}^{eV}d\epsilon\;\mathcal{T}(\epsilon)[1-\mathcal{T}(\epsilon)]\\&=\frac{2e^2}{h}\bigg\{\Gamma\bigg[\arctan\bigg(\frac{eV-\varepsilon_0}{\Gamma}\bigg)+\arctan\bigg(\frac{\varepsilon_0}{\Gamma}\bigg)\bigg]+\\&-\frac{\Gamma^2\varepsilon_0}{\varepsilon_0^2+\Gamma^2}-\frac{\Gamma^2(eV-\varepsilon_0)}{(eV-\varepsilon_0)^2+\Gamma^2}\bigg\}=-S_{LR}(0,V),
\end{split}
\end{equation}
In the linear regime (i.e. at low voltage $V$), the noise can be approximated as
\begin{equation}
\begin{split}
    S_{LL}(0,V)&=\frac{4e^2}{h}eV\mathcal{T}(0)[1-\mathcal{T}(0)]\\&=\frac{4e^2}{h}eV\frac{\Gamma^2}{\varepsilon^2_0+\Gamma^2}\bigg[1-\frac{\Gamma}{\varepsilon^2_0+\Gamma^2}\bigg],
\end{split}
\end{equation}
and thus it disappears for a symmetric junction at the resonance $\varepsilon_0=0$. In the tunneling limit (small coupling $\Gamma$), one gets the well-known \emph{Schottky formula} \cite{bib:schottky}, $S_{LL}=2GeV=2eI$, which is used in different situations to determine the charge of the carriers.

\begin{figure}
  \includegraphics[width=\linewidth]{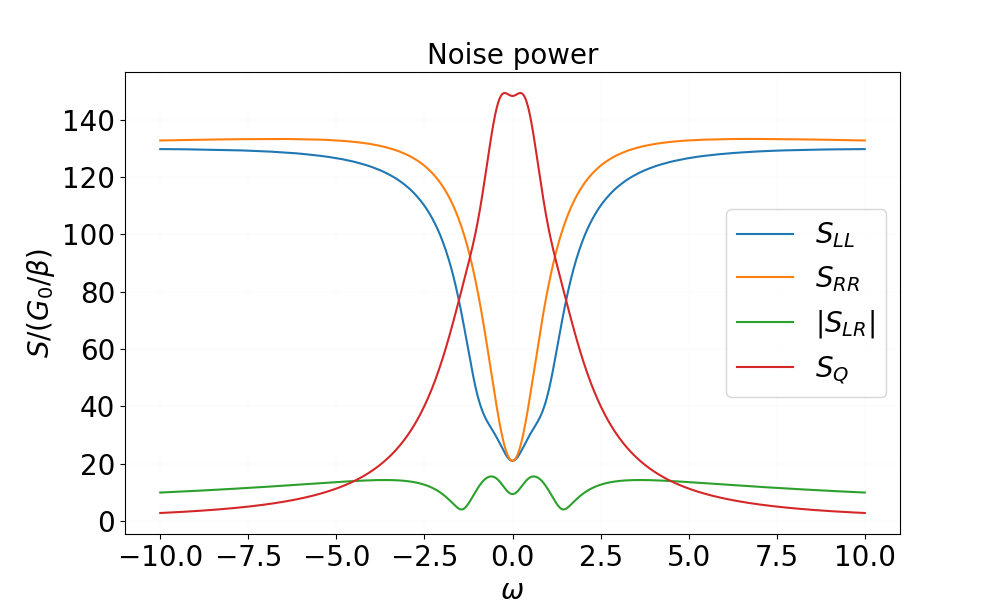}
  \caption{Current noise power elements $S_{LL}$, $S_{RR}$, $|S_{LR}|$, and charge noise power $S_Q$ (in units of $G_0/\beta$), as a function of frequency $\omega$ at the resonance $\varepsilon_0=0$ at fixed temperature ($\beta=10$) and potential ($eV=1$), with $\Gamma=0.5$.}
  \label{fig:8}
\end{figure}

\begin{figure}
\includegraphics[width=\linewidth]{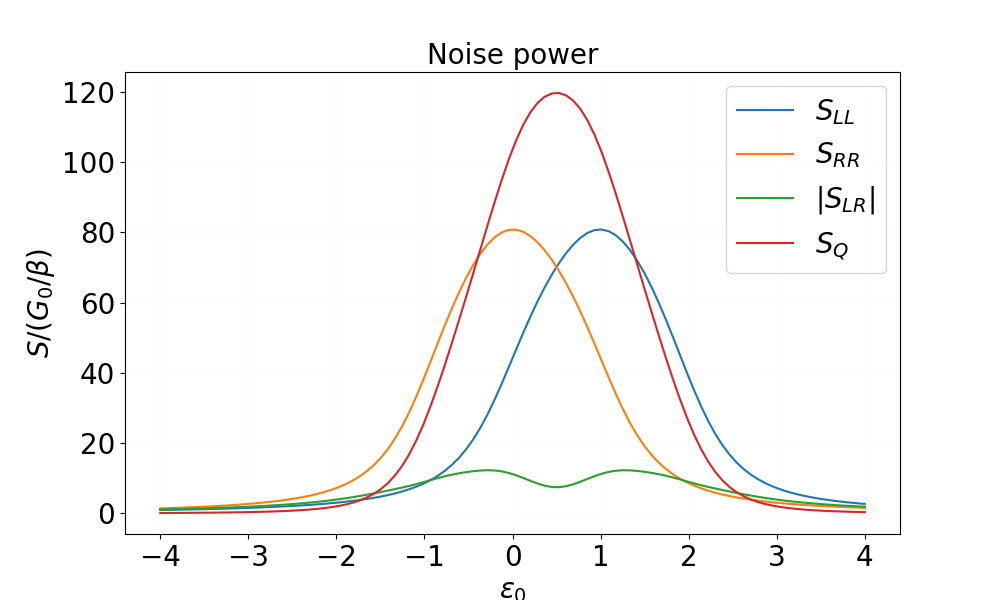}
  \caption{Current noise power elements $S_{LL}$, $S_{RR}$, $|S_{LR}|$, and charge noise power $S_Q$ (in units of $G_0/\beta$), as a function of dot's energy $\varepsilon_0$ at fixed temperature ($\beta=10$), potential ($eV=1$), and frequency ($\omega=1$), with $\Gamma=0.5$.}
  \label{fig:9}
\end{figure}

\begin{figure}
    \includegraphics[width=\linewidth]{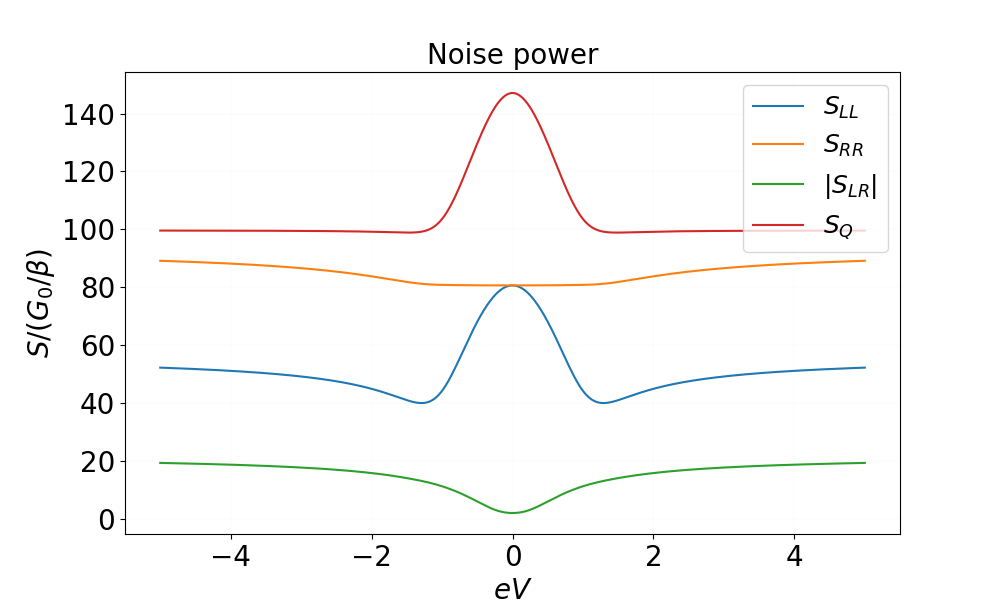}
  \caption{Current noise power elements $S_{LL}$, $S_{RR}$, $|S_{LR}|$, and charge noise power $S_Q$ (in units of $G_0/\beta$) as a function of potential $eV$ at the resonance $\varepsilon_0=0$ at fixed temperature ($\beta=10$) and frequency $\omega=1$, with $\Gamma=0.5$.}
  \label{fig:10}
\end{figure}

We show the current noise power elements, $S_{LL}$, $S_{RR}$, $|S_{LR}|$, and the charge noise power $S_Q$, as a function of the frequency $\omega$ at the resonance $\varepsilon_0=0$ and fixed potential $eV$ in Fig. \ref{fig:8}. We see that the charge noise shows a maximum at zero frequency, while the auto-correlators $S_{LL}$ and $S_{RR}$ have an equal minimum. In Fig. \ref{fig:9} we plot the noises as a function of the dot's energy $\varepsilon_0$ at fixed $\omega$ and $eV$. We see that the auto-correlator $S_{aa}$ shows a maximum at $\varepsilon_0=\mu_a$, while for the charge noise $S_Q$ it is located at the center $(\mu_L+\mu_R)/2\equiv eV/2$. Finally, in Fig. \ref{fig:10} are reported the noises in function of the potential $eV$ at the resonance $\varepsilon_0=0$ and fixed $\omega$, in which we see that the left and right auto-correlators are generally different except at equilibrium, namely $V=0$, where the noise is purely thermal. These results confirm that in the finite-frequency ($\omega\not=0$) and non-equilibrium ($eV\not=0$) conditions the left and right noises, $S_{LL}$ and $S_{RR}$, are generally different, but there is an exception when $\varepsilon_0=(\mu_L+\mu_R)/2$, namely when the energy corresponds to the mean chemical potential of the two leads. On the other hand, these two quantities always coincide when zero-frequency ($\omega=0$) or equilibrium ($V=0$) are considered
{\color{black} \cite{bib:zamoum}}. Regarding the cross-correlator $S_{LR}$, it seems that there are no particular values of frequency, potential or dot's energy for which the quantum correlations between the current in the left and right leads are particularly strong. This is because the parameter of correlation between the two leads is mainly determined by the coupling $\Gamma$, which is kept fixed. 


\section{Two-level QD}
\label{bib:two}
We now consider the case of a non-interacting dot with two discrete levels $\varepsilon_1$, $\varepsilon_2$ coupled to two metallic leads, with $k$-independent tunneling matrix elements $W_{L1}$, $W_{L2}$ and $W_{R1}$, $W_{R2}$. Now there are four Green's functions in Keldysh space, $\hat{G}_{nm}$, where $n,m=1,2$. By using the \emph{Keldysh-Dyson eqution} (\ref{eq:Keldysh-Dyson}), the retarded Green's functions of the coupled dot can be written as
\begin{equation}
\begin{split}
G^{R}_{11}(\omega)&=\frac{\omega-\varepsilon_2-|W_{L2}|^2g^R_L(\omega-eV)-|W_{R2}|^2g^R_R(\omega)}{D^R(\omega,V)},\\
G^{R}_{12}(\omega)&=\frac{W^*_{L1}W_{L2}g_L^R(\omega-eV)+W^*_{R1}W_{R2}g_R^R(\omega)}{D^R(\omega,V)},\\ 
G^{R}_{21}(\omega)&=\frac{W^*_{L2}W_{L1}g_L^R(\omega-eV)+W^*_{R2}W_{R1}g_R^R(\omega)}{D^R(\omega,V)},\\
G^{R}_{22}(\omega)&=\frac{\omega-\varepsilon_1-|W_{L1}|^2g^R_L(\omega-eV)-|W_{R1}|^2g^R_R(\omega)}{D^R(\omega,V)},
\end{split} 
\end{equation} 
with the denominator
\begin{equation}
\begin{split}
    D^R=&\big[\omega-\varepsilon_1-|W_{L1}|^2g^R_L-|W_{R1}|^2g^R_R\big]\times\\&\big[\omega-\varepsilon_2-|W_{L2}|^2g^R_L-|W_{R2}|^2g^R_R\big]+\\-&|W_L1|^2|W_L2|^2[g^R_L]^2-|W_R1|^2|W_R2|^2[g^R_R]^2+\\-&2\Re[W^*_{L1}W_{L2}W_{R1}W^*_{R2}]g_L^Rg_R^R,
\end{split}
\end{equation}
where we dropped the frequency and voltage dependence for convenience. The Keldysh Green's functions are given by
\begin{equation}
\begin{split}
    G^K_{nm}(\omega)&=\sum_{ij}G^R_{ni}(\omega)\Sigma^K_{ij}(\omega)G^A_{jm}(\omega),\\
\Sigma^K_{nm}(\omega)&=-2\pi i\bigg[F_L(\omega-eV)W^*_{Ln}W_{Lm}\nu_L(\omega-eV)+\\&+F_R(\omega)W^*_{Rn}W_{Rm}\nu_R(\omega)\bigg],
\end{split}
\end{equation}
for $n,m=1,2$. 

We work again in the wide band limit, namely $g^R_a=-i\pi\nu_a$ with complex couplings $\Gamma^a_{nm}=2\pi W^*_{an}W_{am}\nu_a=(\Gamma^a_{mn})^*$, such that $\Gamma^a_{11}\Gamma_{22}^a=|\Gamma_{12}^a|^2$, and we also consider a symmetric junction, $\Gamma^L_{nm}=\Gamma^R_{nm}=\Gamma_{nm}$, for simplicity. By substituting $G^{R(A)}_{nm}$ and $G^K_{nm}$ expressions in the Meir-Wingreen formula (\ref{eq:meir-wingreen}), one still gets an expression for the average current in the Landauer form (\ref{eq:genland2}), with frequency independent transmission coefficient $\mathcal{T}(\omega)$ which now reads
\begin{equation}
    \begin{split}
        \mathcal{T}(\omega)&=\frac{[\Gamma_{11}(\omega-\varepsilon_2)+\Gamma_{22}(\omega-\varepsilon_1)]^2}{[(\omega-\varepsilon_1)(\omega-\varepsilon_2)]^2+[\Gamma_{11}(\omega-\varepsilon_2)+\Gamma_{22}(\omega-\varepsilon_1)]^2}\\&=\frac{[\Gamma(\omega-\varepsilon)-\Delta\varepsilon\Delta\Gamma/2]^2}{[(\omega-\varepsilon)^2-(\Delta\varepsilon)^2/4]^2+[\Gamma(\omega-\varepsilon)-\Delta\varepsilon\Delta\Gamma/2]^2},
    \end{split}
    \label{Tau}
\end{equation}
where in the second equality we introduced the notation $\varepsilon=(\varepsilon_1+\varepsilon_2)/2$, $\Delta\varepsilon=\varepsilon_2-\varepsilon_1$ for the dot energies and $\Gamma=\Gamma_{11}+\Gamma_{22}$, $\Delta\Gamma=\Gamma_{11}-\Gamma_{22}$ for the couplings. Notice that, if the system is degenerate, namely $\Delta\varepsilon=0$, one recovers the transmission coefficient of a single-level QD with effective coupling $\Gamma=\Gamma_{11}+\Gamma_{22}$.

\begin{figure}
  \includegraphics[width=\linewidth]{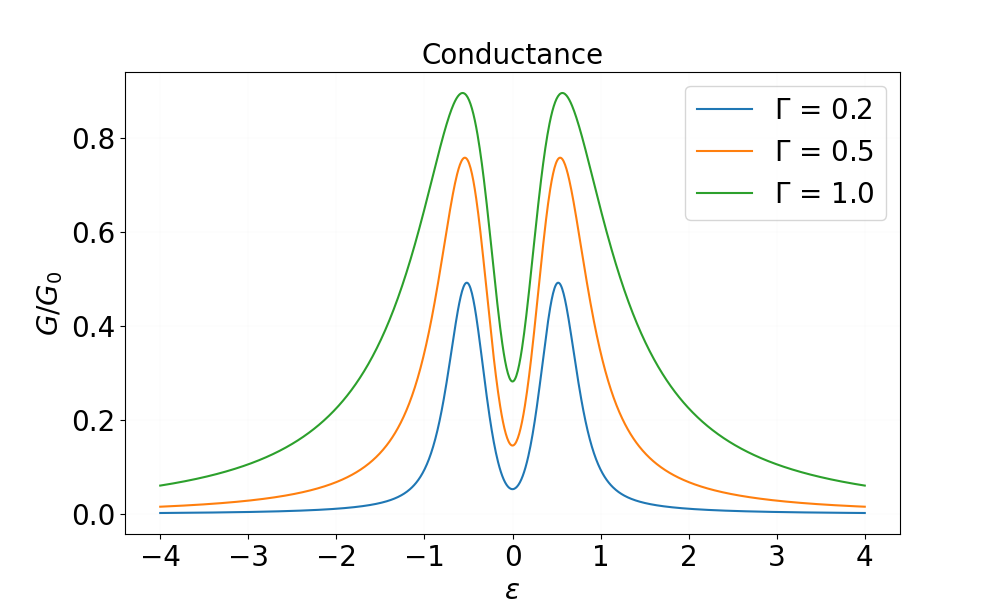}
  \caption{Linear conductance $G$ in units of the conductance quantum $G_0=2e^2/h$ as a function of the mean dot's energy $\varepsilon=(\varepsilon_1+\varepsilon_2)/2$ at fixed temperature ($\beta=10$) for different values of the coupling $\Gamma=\Gamma_{11}+\Gamma_{22}$ and $\Delta\Gamma=\Gamma_{11}-\Gamma_{22}=0$ in a non-degenerate two-level QD ($\Delta\varepsilon=\varepsilon_2-\varepsilon_1=1$).}
  \label{fig:4}
  \includegraphics[width=\linewidth]{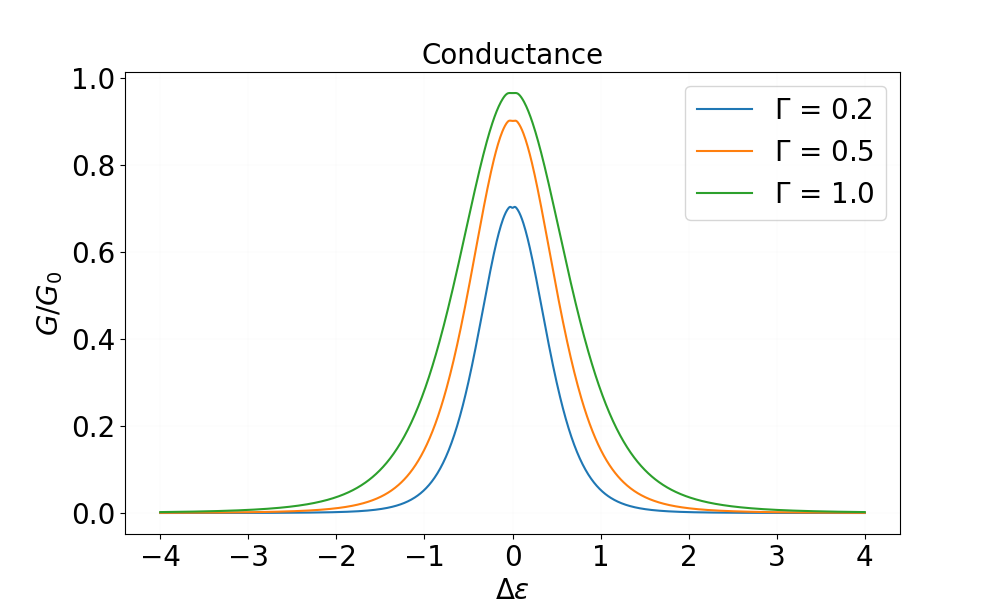}
  \caption{Linear conductance $G$ in units of the conductance quantum $G_0=2e^2/h$ as a function of the energy distance between the dot's levels $\Delta\varepsilon=\varepsilon_2-\varepsilon_1$ at fixed temperature ($\beta=10$) with average energy at the Fermi level $\varepsilon=(\varepsilon_1+\varepsilon_2)/2=0$ for different values of the coupling $\Gamma{=\Gamma_{11}+\Gamma_{22}}$ {and $\Delta\Gamma=\Gamma_{11}-\Gamma_{22}=0$}.}
  \label{fig:5}
\end{figure}

In the linear regime the conductance reads
\begin{equation}
\begin{split}
    G=-\frac{2e^2}{h}\int_{-\infty}^{\infty}d\omega\;\frac{df_L}{d\omega}(\omega)\mathcal{T}(\omega),
\end{split}
\label{G2!}
\end{equation}
which, at zero-temperature, has the analytical form
\begin{equation}
\begin{split}
    G&=\frac{2e^2}{h}\frac{(\Gamma_{11}\varepsilon_2+\Gamma_{22}\varepsilon_1)^2}{\varepsilon^2_1\varepsilon^2_2+(\Gamma_{11}\varepsilon_2+\Gamma_{22}\varepsilon_1)^2}\\&={
    \frac{2e^2}{h}\frac{(\Gamma\varepsilon+\Delta\varepsilon\Delta\Gamma/2)^2}{(\varepsilon^2-\Delta\varepsilon^2/4)^2+(\Gamma\varepsilon+\Delta\varepsilon\Delta\Gamma/2)^2}}.
    \label{GT0}
\end{split}
\end{equation}
We see that the symmetric conductance is maximal if one of the two levels is resonant (i.e. at the Fermi energy) and is equal to the conductance quantum $G_0=2e^2/h$ in the case of zero-temperature, irrespective of the strength of the couplings to the electrodes. We notice that one can also find an \emph{anti-resonance} condition, namely $\varepsilon_1=-(\Gamma_{11}/\Gamma_{22})\varepsilon_2$, for which the conductance is minimal, i.e. $G=0$ in the zero-temperature case. Interestingly one can see from Eq. (\ref{GT0}) that, starting from the zero-temperature degeneracy $\varepsilon=0$, $\Delta\varepsilon=0$ with $\Delta\Gamma=0$ and then shifting symmetrically the two levels $\varepsilon=0$, $\Delta\varepsilon\not=0$, one passes from $G=G_0$ to $G=0$ discontinuously. 
{\color{black} On the other hand, at finite temperature one can easily show numerically from Eq.~(\ref{G2!}) that the full width at half maximum of $G$ as a function of $\Delta \varepsilon$ vanishes approaching zero temperature as $\propto T^{\alpha}$ with $\alpha\approx 0.5$}. 
In other words, one can block the current in the system by a simple symmetrical removal of the degeneracy. At finite temperature, the results for the linear conductance as a function of the mean dot's energy $\varepsilon=(\varepsilon_1+\varepsilon_2)/2$ and the energy distance between the levels $\Delta\varepsilon=\varepsilon_2-\varepsilon_1$  are reported in Fig. \ref{fig:4} and Fig. \ref{fig:5}, respectively, for different values of the coupling $\Gamma=\Gamma_{11}+\Gamma_{22}$ {with $\Delta\Gamma=\Gamma_{11}-\Gamma_{22}=0$}. One can clearly see the presence of the anti-resonance which minimizes the conductance close to zero, that vanishes in the case of degeneracy $\Delta\varepsilon=0$, for which one can obtain a single-peaked conductance as in the case of a single-level QD. Moreover, one can also see that the conductance drops to zero continuously (due to thermal effects) by a symmetrical removal of the degeneracy of the dot when $\varepsilon=0$, allowing to block the current in the system.

\begin{figure}
  \includegraphics[width=\linewidth]{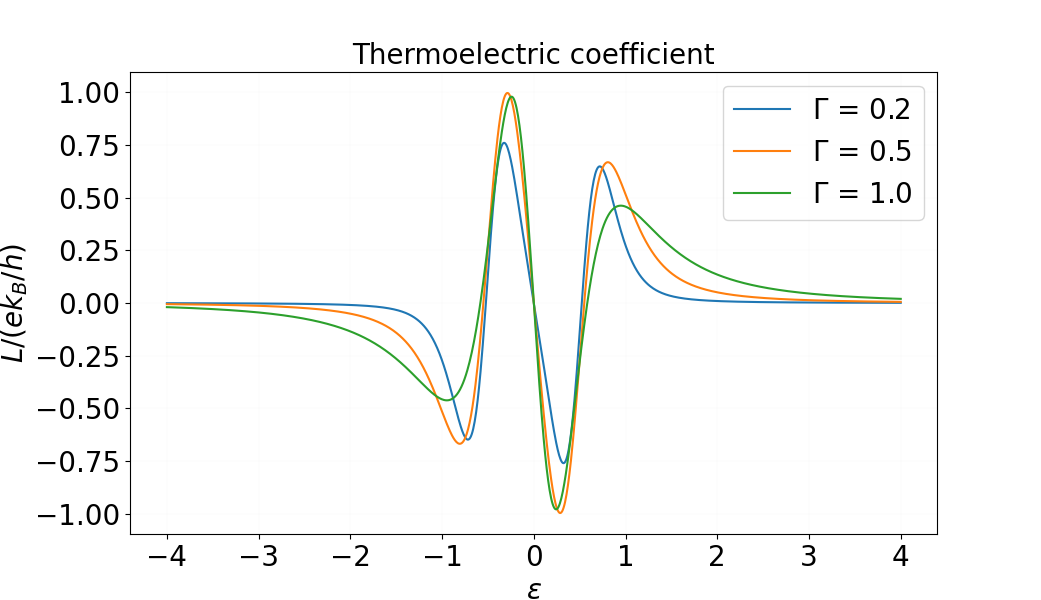}
  \caption{Thermoelectric coefficient $L$ in units of {\color{black}$ek_B/h$} as a function of the mean dot's energy $\varepsilon=(\varepsilon_1+\varepsilon_2)/2$ at fixed temperature ($\beta=10$) for different values of the coupling $\Gamma=\Gamma_{11}+\Gamma_{22}$ {and $\Delta\Gamma=\Gamma_{11}-\Gamma_{22}=0$} in a non-degenerate two-level QD ($\Delta\varepsilon=\varepsilon_2-\varepsilon_1=1$).}
  \label{fig:6}
  \includegraphics[width=\linewidth]{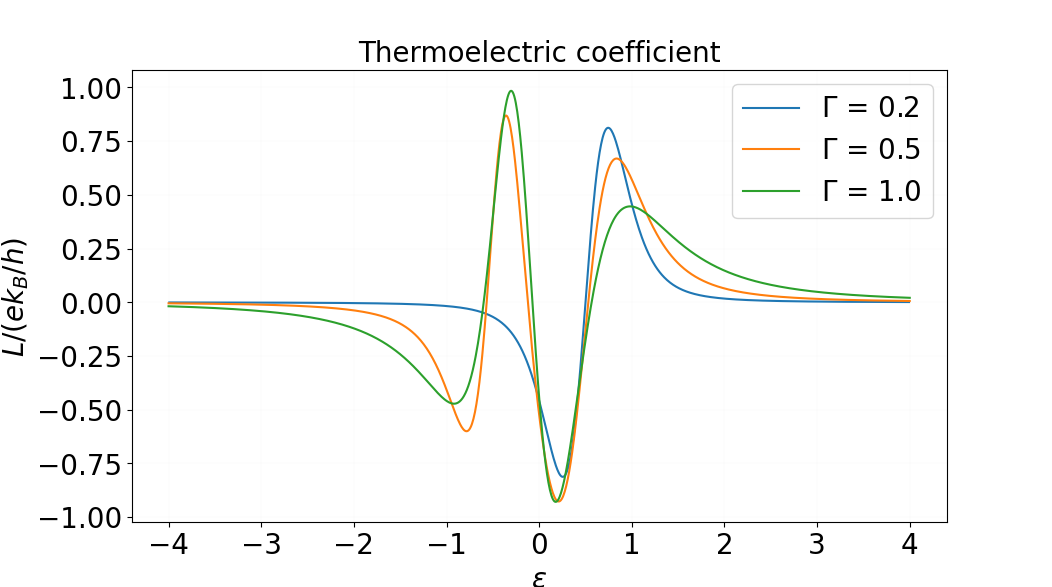}
  \caption{Thermoelectric coefficient $L$ in units of {\color{black}$ek_B/h$} as a function of the mean dot's energy $\varepsilon=(\varepsilon_1+\varepsilon_2)/2$ at fixed temperature ($\beta=10$) for different values of the couplings $\Gamma=\Gamma_{11}+\Gamma_{22}$ and $\Delta\Gamma=\Gamma_{22}-\Gamma_{11}=0.2$ in a non-degenerate two-level QD ($\Delta\varepsilon=\varepsilon_2-\varepsilon_1=1$).}
  \label{fig:7}
\end{figure}

On the other hand, one has the thermoelectric coefficient
\begin{equation}
    L=-\frac{2ek_B}{h}\int_{-\infty}^{\infty}d\omega\,\beta\omega\,\frac{df_L}{d\omega}(\omega)\mathcal{T}(\omega),
\end{equation}
which, as expected, is always zero in the particle-hole symmetric case, namely $\Delta\Gamma=0$ with $\varepsilon=0$, that is a particular case of anti-resonance in the two-level QD. Moreover, similarly to the single-level case, the thermoelectric coefficient $L$ is an odd function in the mean dot's energy $\varepsilon$ when $\Delta\Gamma=0$, as shown in Fig. \ref{fig:6}, which means that one can control the direction of the thermoelectric current by moving the mean energy of the two-level system. 
More generally, when $\Delta\Gamma\not=0$, the magnitude and the sign of the thermoelectric coefficient depends non-trivially on the parameters $\varepsilon_1, \varepsilon_2, \Gamma_{11}, \Gamma_{22}$, as shown in Fig. \ref{fig:7}.\\
{\color{black} Also in this two-level case both in the conductance and in the thermoelectric coefficient the temperature's dependence can be absorbed by rescaling the parameters, 
\begin{eqnarray}
\label{G2}
    G=\frac{2e^2}{h}\int_{-\infty}^{\infty}dx\;\frac{e^x}{(1+e^x)^2}\,\widetilde{\cal{T}}(x),\\
      L=\frac{2ek_B}{h}\int_{-\infty}^{\infty}dx\;\frac{x\,e^x}{(1+e^x)^2}\,\widetilde{\cal{T}}(x),
      \label{L2}
\end{eqnarray}
where $\widetilde{\cal{T}}(x)\equiv\mathcal{T}(x, \beta\Gamma_{11},\beta\Gamma_{22},\beta\varepsilon_1,\beta\varepsilon_2)$, namely where in Eq.~(\ref{Tau}) we resale $\Gamma_{ss}\rightarrow \beta\Gamma_{ss}$ and $\varepsilon_s\rightarrow \beta\varepsilon_s$.

Let us consider for simplicity $\Delta \Gamma=0$. 
In the weak coupling limit, for $\Delta\epsilon\gg \Gamma$ and $\beta\Gamma\ll 1$, we can approximate
\begin{equation}
\label{appTau}
\tilde{\cal T}(x)\approx \sum_{s=1,2}\frac{(\beta\Gamma/2)^2}{(x-\beta \varepsilon_s)^2+(\beta\Gamma/2)^2}\approx \frac{\pi}{2}\beta\Gamma\sum_{s=1,2}\delta(x-\beta\varepsilon_s)
\end{equation}
getting simple analytical forms for $G$ and $L$ when replacing Eq.~(\ref{appTau}) in Eqs.~(\ref{G2}), (\ref{L2}), 
which vanish linearly upon decreasing $\beta\Gamma$. 
In the strong coupling limit, $\beta\Gamma\gg 1$, instead, $G$ reaches trivially its maximum value, $G_0$, while $L$ is again suppressed.\\ 
As done for the single-level dot, we can find the main peaks for $L$ from Eq.~(\ref{L2}). For $\Delta \Gamma=0$, the peaks occur at  
$\beta\Gamma\approx 3.7, \;
\beta\varepsilon\approx \pm 3.1, \;
\Delta\varepsilon \approx 11.5$.
The maximum value of $L$, in this case, is
\begin{equation}
\label{L2max}
L_{max} \approx \frac{e k_B}{h}.
\end{equation}
In physical situations $\Gamma_{11}$ and $\Gamma_{22}$ do not differs very much therefore the approximated value of the upper limit,  Eq.~(\ref{L2max}), can be approximately valid also for finite $\Delta\Gamma$. In the extreme situation where e.g. $\Gamma_{11}\neq 0$ and $\Gamma_{22}=0$ we recover the result of the single level dot. 
}

\section{Conclusions}
\label{sec:conclusion}
In this work, we have employed the Keldysh field theory approach to compute the current and noise power matrix in a quantum dot system interfaced with metallic reservoirs. Our findings underscore the efficacy of this method in swiftly retrieving established results from the literature while providing a systematic framework for tackling quantum transport phenomena under non-equilibrium conditions. Through detailed analyses of electric and thermoelectric properties, we have elucidated intriguing behaviors inherent in single-level and two-level non-interacting quantum dots.
 
We recover, for instance, how the thermoelectric coefficient can be manipulated by modulating the mean energy of the quantum dot relatively to the Fermi energy of the leads. 
One of the notable outcomes of our investigation is the finding of the optimal conditions to maximize the thermoelectric transport, getting a universal upper limit for the thermoelectric coefficient. These observations not only highlight the tunability of thermoelectric properties in nanostructured systems but also {underscore} the significance of quantum dot platforms in advancing our understanding of energy conversion processes at the nanoscale.

Furthermore, our examination of two-level quantum dots reveals a remarkable discontinuity in the zero-temperature linear conductance, 
and, therefore, a sudden switching-off of the electric current,  
following a symmetrical removal of degeneracy at the Fermi energy. This observation underscores the intricate interplay between quantum coherence and electronic correlations in nanostructured systems, offering valuable insights into the underlying mechanisms governing transport phenomena.

In summary, our study contributes to the burgeoning field of electronic and thermoelectric transport in quantum dot-based junctions, shedding light on fundamental aspects of nanoscale transport and paving the way for future advancements in the design and optimization of nanostructured devices for diverse applications ranging from quantum information processing to efficient energy harvesting.

\subsection*{Acknowledgements}
The authors acknowledge financial support from the project BIRD 2021 ``Correlations, dynamics and topology in long-range quantum systems'' of the Department of Physics and Astronomy, University of Padova,
and from the European Union-Next Generation EU within the ``National Center for HPC, Big Data and Quantum Computing'' (Project No. CN00000013, CN1 Spoke 10 - Quantum Computing).

\appendix*
\section{Charge Fluctuations}\label{app:A}

\begin{figure}
  \includegraphics[width=\linewidth]{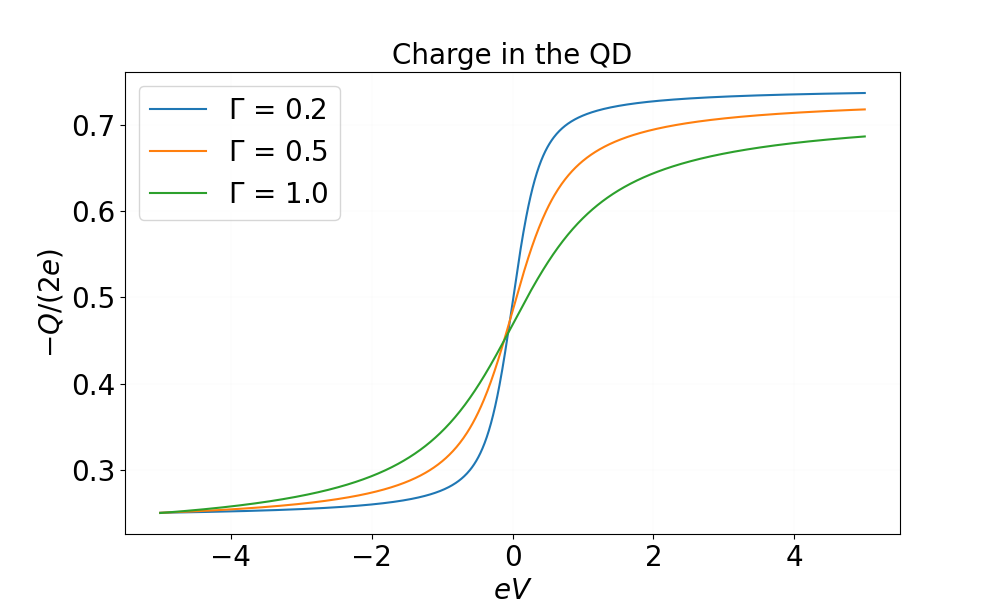}
  \caption{Average charge $Q$ in the dot in units of $2e$ as a function of the potential $eV$ at fixed temperature ($\beta=10$) and dot's energy ($\varepsilon_0=0$), for different values of the coupling $\Gamma$. 
}
  \label{fig:3}
\end{figure}

We want to show the relation (\ref{eq:chargefluc}) between the charge noise in the QD and the current noise matrix. This result has been already obtained by Zamoum et al. \cite{bib:zamoum} within the operatorial Keldysh formalism, but here we show how it can be derived straightforwardly within our field-theoretical approach. The charge operator of the quantum dot is defined as
\begin{equation}
    \hat{Q}(t)=-e\hat{N}(t)=-e\sum_{n}\sum_\sigma\hat{d}^\dagger_{n\sigma}(t)\hat{d}_{n\sigma}(t),
\end{equation}
generated by the quantum source term in the action
\begin{equation}
\begin{split}
    S_B&=-2\int_{-\infty}^\infty dt\; B^q(t)Q^{cl}(t)\\&=e\sum_n\sum_{\sigma}\int_{-\infty}^\infty dt\;B^q(t)\Bar{\Psi}^T_{n\sigma}(t)\hat{\tau}_x\Psi_{n\sigma}(t),
\end{split}
\end{equation}
where
\begin{equation}
    Q^{cl}=\frac{1}{2}(Q^++Q^-)\equiv-\frac{e}{2}\sum_{n}\sum_\sigma(\bar\psi_{n\sigma}^+\psi^+_{n\sigma}+\bar\psi_{n\sigma}^-\psi^-_{n\sigma})
\end{equation}
in such a way that the expectation value of the charge (up to constant term, due to the ambiguity of the time-ordering operator at equal time \cite{bib:kamenev}) is given by
\begin{equation}
\label{eq:avQ}
    Q=\langle Q^{cl}(t)\rangle=\frac{i}{2}\frac{\delta}{\delta B^q(t)}\mathcal{Z}[B^q]\bigg|_{B^q=0}.
\end{equation}
By considering the assumption (\ref{eq:quadraticdot}) of a quadratic action for the QD, one can get the the Keldysh generating functional
\begin{equation}
\begin{split}
    \mathcal{Z}[B^q]&=\Tr \ln \bigg[1+e\hat{G}\hat{B}\bigg]\\&\simeq\Tr\bigg[e\hat{G}\hat{B}^q-\frac{e^2}{2}\hat{G}\hat{B}^q\hat{G}\hat{B}^q\bigg]+O([B^q]^3),
\end{split}
\end{equation}
where $\hat{B}^q=B^q\hat{\tau}_x$, and the trace includes also the sum over the dot's levels. By taking the derivative in (\ref{eq:avQ}), we get the average charge
\begin{equation}
    Q=i\frac{e}{h}\sum_{nm}\int_{-\infty}^{\infty}d\omega\; G_{nm}^K(\omega),
\end{equation}
which for a single-level non-interacting QD in a symmetric junction reads
\begin{equation}
    Q=-2e\int_{-\infty}^{\infty}d\omega \big[f_L(\omega-eV)+f_R(\omega)\big]\frac{\mathcal{T}(\omega,V)}{2\pi\Gamma},
\end{equation}
where we already removed the constant term which do not provides physical meaning. In Fig. \ref{fig:3} we plot the average charge $Q$ in the QD as a function of the electric potential $eV$ for a symmetric junction at equal temperature $\beta$ and different couplings $\Gamma$.

Similarly to current noise power, one can define the \emph{charge noise}, defined as
\begin{equation}
\begin{split}
S_Q(\omega)&=\int_{-\infty}^{\infty}dt\;e^{i\omega t}\langle\delta\hat{Q}(t)\delta\hat{Q}(0)+\delta\hat{Q}(0)\delta\hat{Q}(t)\rangle\\&=\frac{i^2}{2}\frac{\delta^2}{\delta B^q(\omega)\delta B^q(-\omega)}\ln \mathcal{Z}[B^q]\bigg|_{B^q=0},
\end{split}
\end{equation}
where $\delta\hat{Q}(t)=\hat{Q}(t)-Q$. By taking the second derivative, we get
\begin{equation}
\begin{split}
    S_Q(\omega)&=\frac{e^2}{h}\sum_{nm}\int_{-\infty}^{\infty} d\epsilon\; \mbox{tr}\bigg[\hat{G}_{nm}(\epsilon_+)\hat{\tau}_x\hat{G}_{mn}(\epsilon_-)\hat{\tau}_x\bigg]\\&=\frac{e^2}{h}\sum_{nm}\int_{-\infty}^{\infty} d\epsilon\;\bigg[G^K_{nm}(\epsilon_+)G^K_{mn}(\epsilon_-)+\\& G^R_{nm}(\epsilon_+)G^A_{mn}(\epsilon_-)+G^A_{nm}(\epsilon_+)G^R_{mn}(\epsilon_-)\bigg],
\end{split}
\end{equation} 
which for a non-interacting single-level dot reads
\begin{equation}
    S_Q(\omega)=\frac{e^2}{h}\int_{-\infty}^{\infty} d\epsilon\; \frac{\mathcal{T}(\epsilon_+)\mathcal{T}(\epsilon_-)}{\Gamma^2}\sum_{ab=L,R}B_{ab}(\epsilon,\omega).
\end{equation}
The plots for the charge noise power are reported in Figg. \ref{fig:8}-\ref{fig:10}. In our case, one can check the simple relation
\begin{equation}
    |t(\epsilon+)-t(\epsilon_-)|^2=\frac{\omega^2}{\Gamma^2}\mathcal{T}(\epsilon_+)\mathcal{T}(\epsilon_-),
\end{equation}
which leads to the equality that we wanted to prove
\begin{equation}
    \sum_{ab=L,R}S_{ab}(\omega)=\omega^2S_Q(\omega).
\end{equation}

\onecolumngrid

\bigskip 

\end{document}